# High Capacity Hydrogen Storage on Zirconium decorated γ-graphyne: A systematic first-principles study

*Mukesh Singh[a], Alok Shukla[a], Brahmananda Chakraborty[b,c,\*]*,

[a]Department of Physics, Indian Institute of Technology Bombay, Powai, Mumbai 400076, India

[b]High Pressure and Synchrotron Radiation Physics Division, Bhabha Atomic Research Centre, Trombay, Mumbai, India

[c]Homi Bhabha National Institute, Mumbai, India

**Abstract:** In this work, we investigate the hydrogen-storage properties of Zr-decorated γ-graphyne monolayer employing Density Functional Theory (DFT) for green energy storage. We predict that each Zr atom decorated on graphyne sheet (2D) can adsorb up to seven $H_2$ molecules with an average adsorption energy of -0.44 eV/$H_2$, leading to a hydrogen gravimetric density of 7.95 wt%, and desorption temperature of 574 K, particularly suited to fuel-cell applications. Decorated Zr atom strongly attached to graphyne due to charge transfer from Zr to graphyne sheet. Hydrogen molecules adsorb on Zr decorated graphyne with Kubas type of interaction. The 4.05 eV diffusion energy barrier between Zr decorated position, and its neighboring pores may avoid the metal-metal (Zr-Zr) clustering. The stability of Zr+γ-graphyne is confirmed by performing ab-initio molecular dynamics simulations at room temperature and at estimated average desorption temperature. Hence, our calculations show Zr functionalized on γ-graphyne could be a promising solid-state hydrogen storage material.

___________________

*Corresponding author: Dr. Brahmananda Chakraborty
E-mail address: brahma@barc.gov.in
Phone: +91-2225592057

## Introduction

Depleting fossil fuels coupled with the perpetually increasing demand for energy and pollution has forced humankind to search for alternative energy sources. Considering the population and economic growth, Shafiee and Topal have estimated that oil, gas, and coal will diminish in ~92 years [1]. For alternative energy resources, factors like availability of raw materials, manufacturing and transportation costs, the convenience of utilization, efficiency, health and safety, environmental impact, etc., must be considered [2]. The US Department of Energy has categorized all available alternative sources and specified criteria for finding efficient and widely usable alternative sources of energy [3–9]. Hydrogen energy is a promising alternative, but unfortunately, it does not exist naturally as a primary source of energy. It is a secondary type of energy known as an energy carrier. Major advantages of hydrogen energy over other alternatives are: (a) the only byproducts after its utilization are heat and water, making it environmentally favorable and generating a recyclable process [10]. b) It is abundant in nature as a constituent of water ($H_2O$). c) Its high gravimetric energy density (143 $MJKg^{-1}$) is approximately three times higher than methane, LPG propane, butane, biodiesel, diesel, natural gas, etc. [11]. d) Local hydrogen plants can reduce transportation cost-effectively as fossil fuels are required to transfer first from their natural reserves to the refineries and finally to the places of utilization. Considering the plant's consumption of $CO_2$ from the environment produces not only biomass [12] but also beneficial extract exploits for $H_2$ production and nanoparticle synthesis [13]. Other techniques, including electrolysis and photocatalysis [14], have also been advised for $H_2$ production [15]. Having a low boiling temperature, 20 K [16], and being highly combustive, hydrogen poses many challenges in its storage. Several approaches aimed at hydrogen storage involving mechanical to chemical mechanisms have been employed so far and can be classified as: a) high-pressure gas storage, b) cryogenic liquid hydrogen, c) physically bound hydrogen, d) chemically bound hydrogen, and e) hydrolytic evolution of hydrogen [17]. In terms of applications, hydrogen storage can be categorized into stationary hydrogen storage and onboard hydrogen storage.

Conventional highly compressed gas storage methods require high pressure and a massive tank to store it, whereas liquid hydrogen storage necessitates cryogenic temperature because of the lightest weight of hydrogen atoms. Both of these approaches are too expensive and unsuitable for hydrogen storage onboard. A compact, safe, reliable, and low-budget hydrogen storage device is required in automobiles [18], and one needs to replace conventional mechanical storage devices with solid-state, non-mechanical hydrogen storage devices. For solid-state storage, DoE has specified some criteria for a material to qualify as an effective storage material: a) the binding energy of absorption hydrogen must range between 0.2-0.7 eV, and b) the gravimetric weight percentage of hydrogen storage should be higher than 6.5 [19]. Before the arrival of carbon nanomaterials, various solid-state hydrogen storage methods have been explored, known as conventional solid-state methods and corresponding materials or devices as conventional solid-state materials or devices. These materials can be categorized into four major categories: metal hydrides, metal alloys, zeolites, and metal-organic frameworks (MOFs). Hydrogen creates hydrides with some metals and their alloys, e.g., $LiH_2$, $MgH_2$, $LiAlH_4$, $NaAlH_4$, $LaNiH_6$, $TiFeH_2$, etc., from which solid-state devices can be made.

Under moderate temperatures and pressures, solid-state devices are much safer than gas and liquid-based hydrogen storage setups. The volumetric density of metal hydride (6.5 H atoms/cm$^3$ for MgH$_2$) is higher than that of gas (0.99 H atoms/cm$^3$) or liquid (4.2 H atoms/cm$^3$). Hence, metal hydrides are safe and volumetrically efficient storage for transport applications [20]. Unfortunately, hydrogens in hydrides are chemically bonded, resulting in large binding energies leading to high desorption temperatures. Next, Zeolites are highly crystalline aluminosilicate materials with large anions containing cavities and channels. Due to their microporous structure, they are used as absorbent and catalysts. Some zeolites and silicates have been studied in the past for hydrogen storage [21,22], and it has been found that micropores play a vital role in their hydrogen storage capabilities. Experimentally, it has been shown that many factors such as large surface area, pore-volume, channel size, internal wall topology, etc., affect hydrogen storage on micropores [23]. These factors play a significant role at low temperature and high pressure for effective [24] hydrogen storage. However, the wt% of H$_2$ absorbed by zeolites is very low under ambient conditions. Finally, Metal-organic frameworks (MOFs) are another category of porous nanostructures that combines organic compounds with metal/metal clusters to store hydrogen [25]. But the problem with MOFs is that their gravimetric wt% is much below the target set by DoE at the room temperature.

Carbon nanostructures are highly porous with a large surface area, lightweight which makes them useful for hydrogen storage. On top of that, 2D carbon nanostructures have two sides that make them ideal for storing additional hydrogen. Many carbon allotropes, including carbon molecules, nanotubes, graphene sheets, and other carbon-derived structures, have been explored for hydrogen devices [26,27]. However, for pristine carbon nanomaterials, only weak physisorption contributes to bonding resulting in an inefficient gravimetric weight percentage [28]. For instance, the 0.4 wt% and <0.2 wt% H$_2$ storage on graphene sheets were measured at 77K and ambient temperature, respectively [29]. On functionalizing carbon nanomaterials with metal atoms, wt% of hydrogen storage increases significantly. For instance, metal decorated fullerenes (C$_{60}$) have been reported to enhance H$_2$ gravimetric weight percentage to ~6-9.5 [30–36]. A similar trend also has been noted in metal doped carbon nanotube (CNT) [37–40] and metal doped graphene [41–44]. As far as experiments are concerned, several results have been recorded. For instance, Samantaray et al. have reported H$_2$ storage wt% of ~4.2/4.6 at 298 K for Pd-Co alloy decorated nitrogen/boron doped graphene oxide [45]. Al/Ni/graphene complex can adsorb a maximum H$_2$ up to ~5.7 wt% at 473 K with a desorption efficiency of 96-97% [46]. Mg and its composite, e.g., Mg/MgH$_2$ decorated on Ni doped graphene, show H$_2$ wt% capacity of more than 6.5 [47].

Graphyne and graphdiyne [48,49] have been synthesized and have a mixture of sp$^1$ and sp$^2$ hybridized bonds, which drove our interest in exploring the role of acetylenic linkage to hydrogen storage. In graphyne, more π-orbitals than graphene are available to interact with the decorated metal. Like other pristine carbon nanomaterials, pristine graphyne shows lower hydrogen capability. However, metal decoration on graphyne increases its hydrogen storage wt% [50]. For instance, Alkali (Li), alkaline (Ca), and transition (Sc, Ti) decorated on graphyne nanotube shows hydrogen storage capacities of 4.82, 5.08, 4.88, and 4.76 wt%, respectively [51]. Wu et al. showed that the hydrogen storage gravimetric density of graphyne monolayer decorated by Ca and 6,6,12-graphyne nanotubes decorated by Sc was as high as 5.6 and 5.4 wt%, respectively [52]. The detailed work of Li decorated on graphyne has also been reported [53,54]. However, alkali/alkaline metals bind weakly to carbon nanomaterials, and hydrogens adsorb with low adsorption energy on alkali/alkaline metal decorated graphyne. For instance, Li decorated on graphyne binds Li and adsorb hydrogens with binding and average adsorption energy of -1.35 eV and -0.27 eV/$H_2$, respectively[53]. Li and Na-decorated boron-graphdiyne structures shows absorption energy between 0.21-0.33 eV/$H_2$ [55]. When we decorate transition metals on graphyne, binding of transition metal and adsorption of hydrogens increases. The careful suggestion of Zr transition metal was made because transition metals with less number of d-electrons favours hydrogen storage [56] than other transition metals.

Our DFT simulations predict that each Zr atom can adsorb seven hydrogen molecules resulting 7.95 gravimetric weight percentage. We investigate the bonding mechanism of the metal atom with graphyne and hydrogen molecules sorption on the graphyne+Zr with the help of projected states' density (PDOS) and Bader electronic charge analysis. We used PBE functional and included the dispersion corrections get approximate values of adsorption of $H_2$ to experiments [57]. Furthermore, the DFT simulations are perfomed at absolute zero temperature, we need to check stability of structures at higher temperature and metal-metal (Zr-Zr) clustering. We have investigated higher temperature stablitiy, the possibility of no clustering by performing ab-initio molecular dynamics, and energy barrier calculations for Zr decorated graphyne.

Our article is mainly organized in computational details, results and discussion in 2nd and 3rd sections, respectively. The results and discussion section includes structural properties, interaction mechanism, AIMD, and presence of energy barrier as subsections. Finally, we have concluded the article in the last section.

## 2. Computational methods

The ionic and electronic spin-polarized simulations have been performed with first-principles DFT by using plane-wave-based code VASP [58,59]. Exchange-correlation potential in Local density approximation (LDA) yields large binding energies for the decorated atoms attached to a surface,

So all our calculations employed the generalized-gradient approximation (GGA) [60]. In order to incorporate the weak dispersion forces due to the van der Waals interaction, our calculations have been corrected using Grimmes's DFT-D2 [61]. Ab-initio molecular dynamics simulations (AIMD)[62][63] were also performed to check the stableness of the GY+Zr at higher temperatures in the following two steps. First, we raised the temperature from 0 K to room temperature (300 K), keeping the number of particles, volume, and energy (NVE) of the system constant, and performed the standard molecular dynamics as implemented in VASP. Second, keeping the temperature constant at 300 K with the help of Nose-Hoover thermostats, fixing the number of particles and volume, i.e., in canonical (NVT) ensemble, stability is checked for 5 ps with time steps of 1 fs. The same process of AIMD simulation is repeated for different configurations at different temperatures, as discussed in AIMD calculations section.

In our calculations, we have chosen 2x2x1 supercell of graphyne sheet with 48 atoms and the vacuum of 15 Å in the perpendicular direction. In these calculations, electronic configurations $4d^25s^25p^0$, $2s^22p^2$, and $1s^2$ are employed for valence electrons in the projector-augmented wave (PAW) pseudopotentials of Zr, C, and H, respectively. After ensuring the convergence of the total energy with respect to the kinetic energy (KE) cutoff and Monkhorst-Pack k-point sampling [64] for plane-wave, KE-cutoff and K-points were taken to be 500 eV and 7x7x1, respectively. The energy threshold for the self-consistent convergence is taken $10^{-5}$ eV, while the force threshold for ensuring convergence during geometry optimization is taken 0.01 eV. Conjuagate-gradient (GC) has been used for geometry optimization calculations.

## 3. Result and discussion

### 3.1 Zr doped graphyne: Structure and electronic properties

In geometrically relaxed graphyne, all the six carbon atoms of hexagonal rings are $sp^2$ hybridized, while in triangular rings, six central carbon atoms are $sp^1$, and the remaining six carbon atoms are $sp^2$ hybridized. In Fig. 1(a), $sp^2$-$sp^2$, $sp^2$-$sp^1$, and $sp^1$-$sp^1$ hybridized bond lengths in relaxed γ-graphyne are 1.43 Å, 1.41 Å, and 1.22 Å, respectively, which are consistent with the previously reported values [65]. After decorating Zr at various positions on the graphyne sheet, the midpoints of the hexagon and triangle are recognized as the two most symmetric and stable positions. Symbolically, let us represent Zr doped on the middle of hexagon and triangle with GY+Zr(h) and GY+Zr(t), respectively.

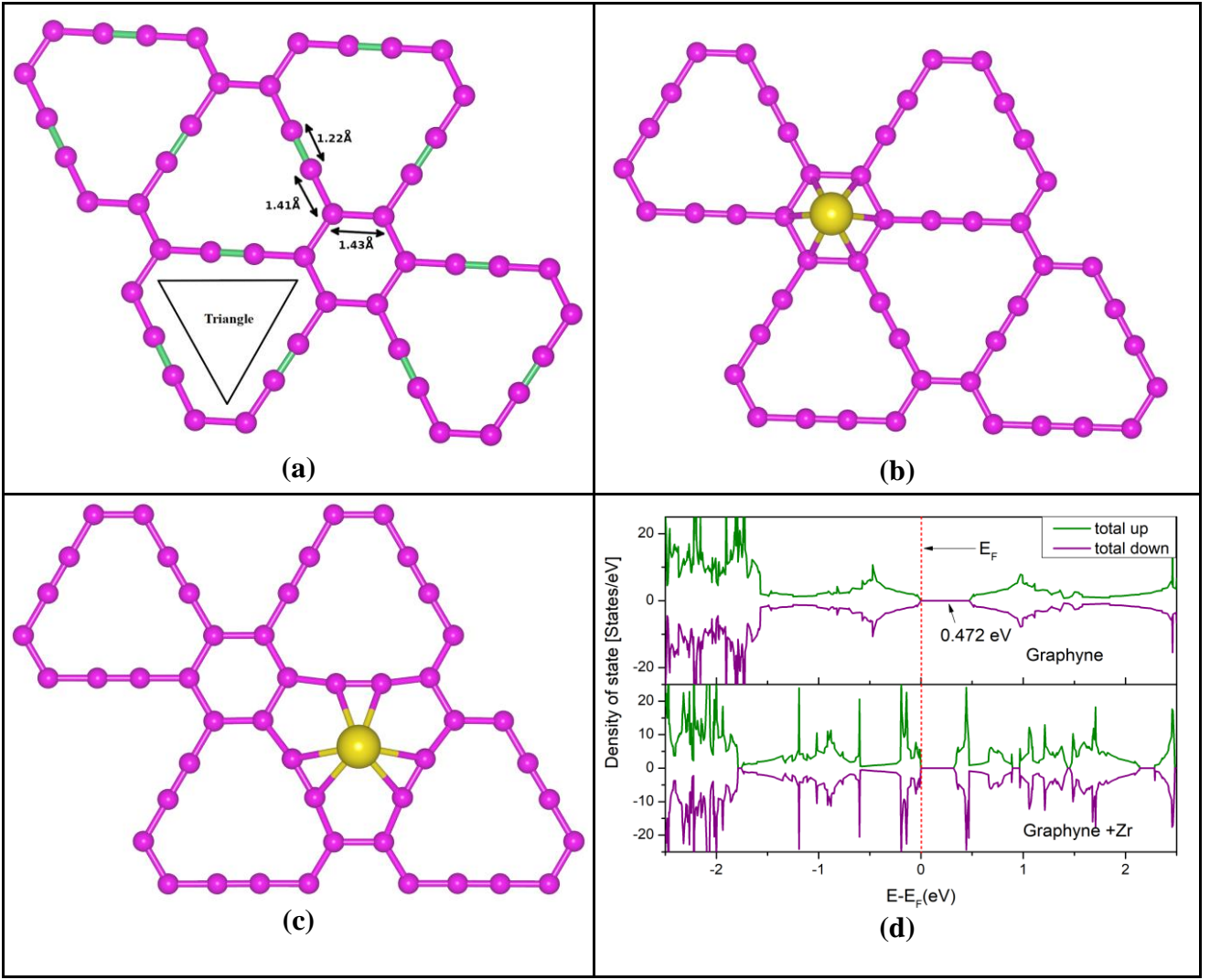

**Fig. 1**: Geometrically relaxed: **(a)** graphyne structure; **(b)** Zr decorated on hexagonal position of graphyne GY+Zr(h); **(c)** Zr decorated on trigonal position of graphyne GY+Zr(t). Pink and golden color spheres represent C and Zr atoms, respectively. Green rods represent acetylenic linkage (sp$^1$-bond). **(d)** The density of states of: GY (upper panel), GY+Zr (lower panel). Fermi energies are equated to zero.

The binding energy of Zr absorbed on graphyne is obtained using:

$$E_b(Zr) = E_{GY+Zr} - E_{GY} - E_{Zr} \qquad (1)$$

Where $E_{GY}$, $E_{Zr}$, and $E_{GY+Zr}$ are DFT calculated ground-state energies of the graphyne monolayer sheet, isolated Zirconium atom, and the Zr doped graphyne monolayer sheet, respectively.

Geometrically relaxed structures of GY+Zr(h) and GY+Zr(t) are plotted in Figs. 1 (b) and (c), respectively. DFT computed binding energies of Zr doped graphyne at hexa- and tri- positions are -3.89 eV, and -5.67 eV, respectively. Both the binding energies are higher than those of Zr doped on graphene (-2.4 eV) [66] and Zr doped on SWCNT [45]. In addition, these binding energies are even higher than any other 3D-transition metal atom decorated on graphyne, except Cr [67]. The presence of extra $sp^1$ hybridized bonds of triangular sites is the main reason behind the stronger binding energy than the hexagonal position. The $sp^2$ hybridized systems have only one π orbital (perpendicular to the plane) per carbon atom available for bonding, while in the $sp^1$ hybridized carbon atoms have two π orbitals (in-plane and perpendicular) that can bind with the metal atom, thus making the bonding stronger at the trigonal position. This is also the reason that the optimized height of Zr atom from the graphyne sheet is 2.26 Å, and 2.31 Å, at the trigonal and the hexagonal positions, respectively, i.e., in the case of tri-position, Zr is closer to the graphyne sheet. Hence acetylenic linkage increases the binding strength of decorated atoms on graphyne sheet.

**3.2 Bonding mechanism of Zr on graphyne:** To understand the interaction of Zr to graphyne sheet qualitatively, we have also analyzed the projected electronic density of states (PDOS) which refers to the contribution of the individual orbitals to the total electronic density of state (TDOS). The total electronic density of Zr decorated graphyne, and pristine graphyne states are plotted in the upper and lower panels of Fig. 1(d), respectively. Our calculated band gap of the pristine graphyne sheet is 0.48 eV, which is consistent with the previous literature value (0.52 eV) [68]. On decorating Zr-atom on graphyne sheet, the density of states plots Fig. 1(d), show the presence symmetric spin polarized states at fermi-energy resulting our system (GY+Zr) metallic and non-magnetic in nature. For the quantitative analysis of the charge distribution in space, we have used "atom in molecules (AIM)" theory of Bader, implemented by Henkelman group [69].

**3.21 PDOS analysis:** Orbital projected density of state's plots show the distribution of electronic states and provide qualitative insights into charge transfers. In Fig. 2, we plotted the projected electrons density of 2p orbital of the nearest C atom for pure GY and the GY+Zr system in the upper panel. We can observe that the charge population at the Fermi level in the Zr doped graphyne is more than that in the pristine graphyne. So there is a charge gain by graphyne when Zr is attached. To understand from where this charge is coming, we plotted the 4d orbital of Zr doped graphyne and isolated Zr atom in the lower panel of Fig. 2. For isolated Zr atom, PDOS are discreet, whereas Zr doped graphyne PDOS are continuous, indicating that Zr is bonded with graphyne sheet. The population of state in isolated Zr is more than that of Zr doped graphyne resulting in a charge loss in the 4d orbital of Zr atom. The PDOS demonstrates that some electrons are transferred from 4d orbital of Zr to 2p orbital of C atoms. So the bonding of Zr at the graphyne sheet because of electron transfer from d-orbital to p-orbitals of Zr and C, respectively, is called p-d hybridization.

**3.22 Bader charge analysis:** For quantitative analysis of electrons transfer, we calculated Bader charge of each atom of our GY+Zr system to analyze the interaction of Zr atom with graphyne in the system. It shows 1.46e charge is reduced from Zr atom and flowed to nearest carbon atoms of the graphyne layer, and redistribute on carbon atoms in GY+Zr. This charge transfer is the reason for the strong interaction between Zr atom and graphyne layer.

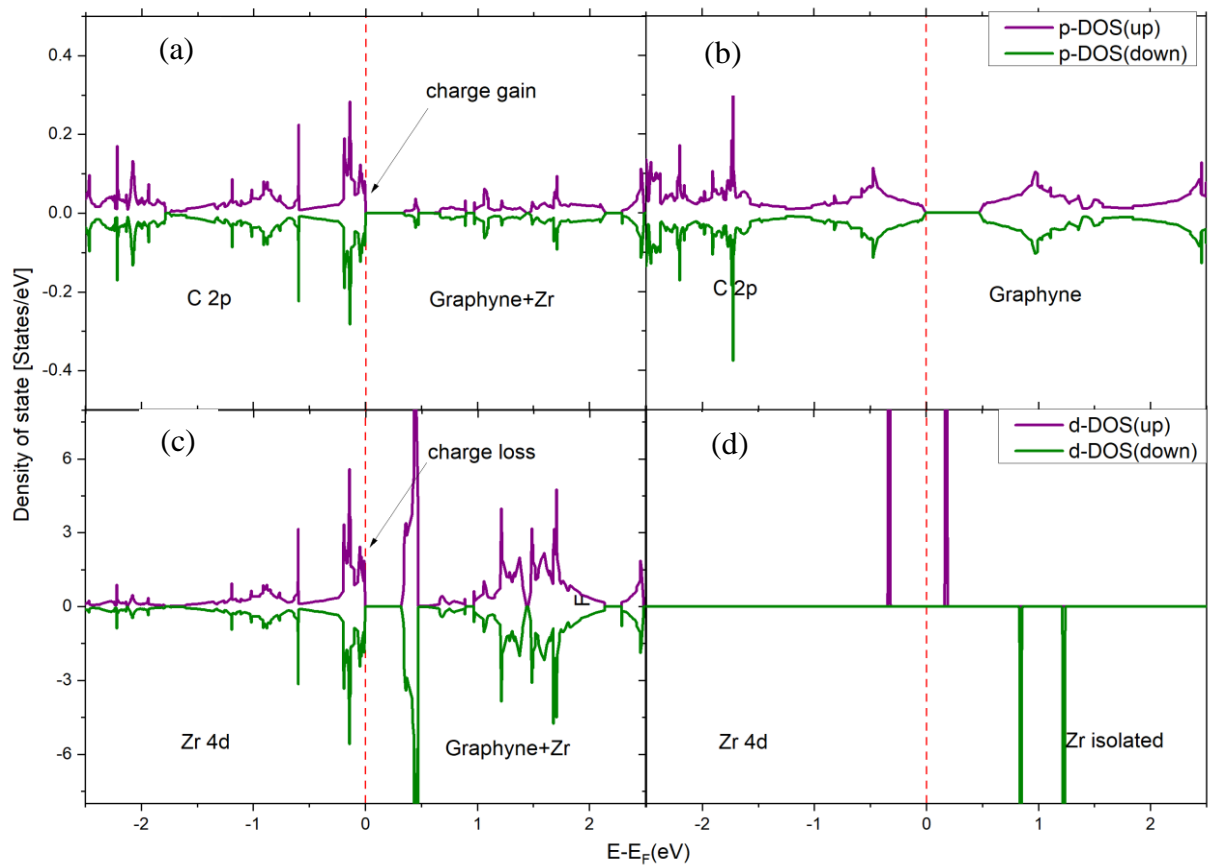

**Fig. 2**:Partial density of states of: **(a)** 2p orbital of C atom in graphyne+Zr; **(b)** 2p orbital of C atom in graphyne; **(c)** 4d orbital of Zr atom in GY+Zr; **(d)** 4d orbital of isolated Zr atom. The enhancement in charge density in (a) as compared to (b), shows that the C atom of graphyne gains charge; The reduction of states in (c) in comparison to (d) shows that Zr atom lost charges, indicating a charge transfer from it to the graphyne sheet.

**3.23 Charge density plots:** From the PDOS and Bader charge analysis, it is clear that Zr atom donates 1.46 electrons to the GY sheet. To visualize the charge transfer in space, the charge density difference, $\rho(GY+Zr)-\rho(GY)$, of GY+Zr and GY is plotted in Fig. 3. The red ring around the Zr-atom shows a depleted charge region, while blue colors on the graphyne sheet show the charge gain region. Hence, after binding of Zr atom on graphyne, most of the transferred charge of Zr atom is distributed among the nearest six carbon atoms of graphyne.

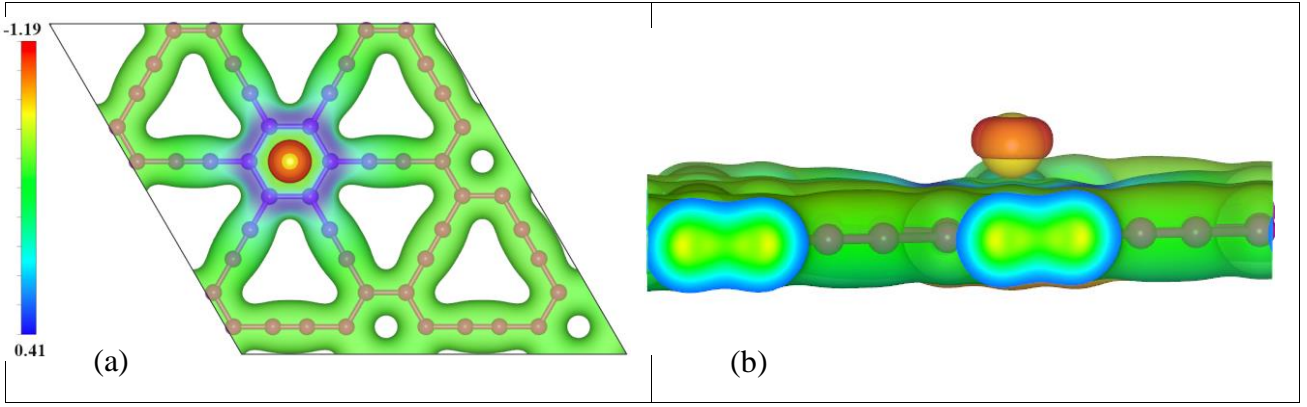

**Fig. 3** The charge density difference of GY+Zr and graphyne, i.e. ρ(GY+Zr)-ρ(GY) with isovalue 0.045e: **(a)** top view **(b)** side view, is plotted. Red and blue regions show charge loss and charge gain regions, respectively.

### 3.3 Hydrogen absorption on Zr decorated graphyne system:

After optimizing the geometry of the GY+Zr system, we started attaching $H_2$ molecules one by one on the Zr atom. First, one $H_2$ molecule was introduced 2.4 Å above Zr atom. The new system (GY+Zr+1$H_2$) was relaxed using GGA-PBE, and corrected the PBE results with Grimme-D2 for dispersion to include weak and long-range contributions of Van der Waals force. In the relaxed system, the first $H_2$ molecule stays 2.40 Å away from Zr atoms, and the bond length of hydrogen molecules stretches from 0.74 Å to 0.77 Å. Using the following formula, we computed the adsorption energy of the first hydrogen molecule:

$$BE_{H_2} = E_{GY+Zr+H_2} - E_{GY+Zr} - E_{H_2} \qquad (2)$$

The adsorption energy of the first $H_2$ molecules is evaluated to -0.45 eV, which is within the DoE specified binding energy range (0.2-0.7 eV) for the suitable hydrogen storage system. Next, we put two more molecules at a distance of 2.4 Å from Zr atoms such that they are at the optimal distance from previously adsorbed $H_2$ and relaxed the new structure.

The mean adsorption energy of second and third hydrogen molecules was computed by using the following equation [70]:

$$BE_{H_2} = \frac{1}{n}\left[E_{GY+Zr+(m+n)H_2} - E_{GY+Zr+mH_2} - nE_{H_2}\right] \qquad (3)$$

The mean adsorption energy of the second and third $H_2$ molecules is evaluated to -0.494 eV. In this configuration, adsorption leads to elongation of the H-H bond of $H_2$ from 0.74 Å to 0.82 Å. To check the maximum capacity, more hydrogen molecules are added systematically until the adsorption energy of $H_2$ molecules stays between the DoE window of 0.2-0.7 eV. Our calculations indicate that up to 7 H$_2$ molecules can be adsorbed on each Zr atom. The average adsorption energies of 4$^{th}$, 5$^{th}$ H$_2$ molecules and 6$^{th}$, 7$^{th}$ H$_2$ are -0.567 eV and -0.262 eV, respectively. The average adsorption energy of our system turns out to be -0.445 eV per H$_2$. The relaxed geometries of 1H$_2$, 3H$_2$, 5H$_2$, and 7H$_2$ doped on the GY+Zr are plotted in Fig. 4, and binding energies are presented in Table 1. In Table 2, our hydrogen storage results are compared with several previously reported theoretical predictions and the experimental works on graphyne-like structures.

**Table 1:** Binding energy of Zirconium on graphyne sheet, adsorption energy of hydrogen molecules on graphyne+Zr and enlargement of H-H bond of adsorbed hydrogen molecules.

| S.N. | System | Van der waals corrected (DFT-D2) Adsorption energy(eV) | Bond length of H-H in H$_2$ molecules(Å) |
|---|---|---|---|
| 1. | GY | -- | -- |
| 2. | GY+Zr | -3.89 | -- |
| 3. | GY+Zr+1H$_2$ | -0.457 | 0.765 |
| 4. | GY+Zr+3H$_2$ | -0.494 | 0.821 |
| 5. | GY+Zr+5H$_2$ | -0.567 | 0.834 |
| 6. | GY+Zr+7H$_2$ | -0.262 | 0.750 |
| Average adsorption energy | | | -0.445 eV |
| Average desorption temperature | | | 574K |

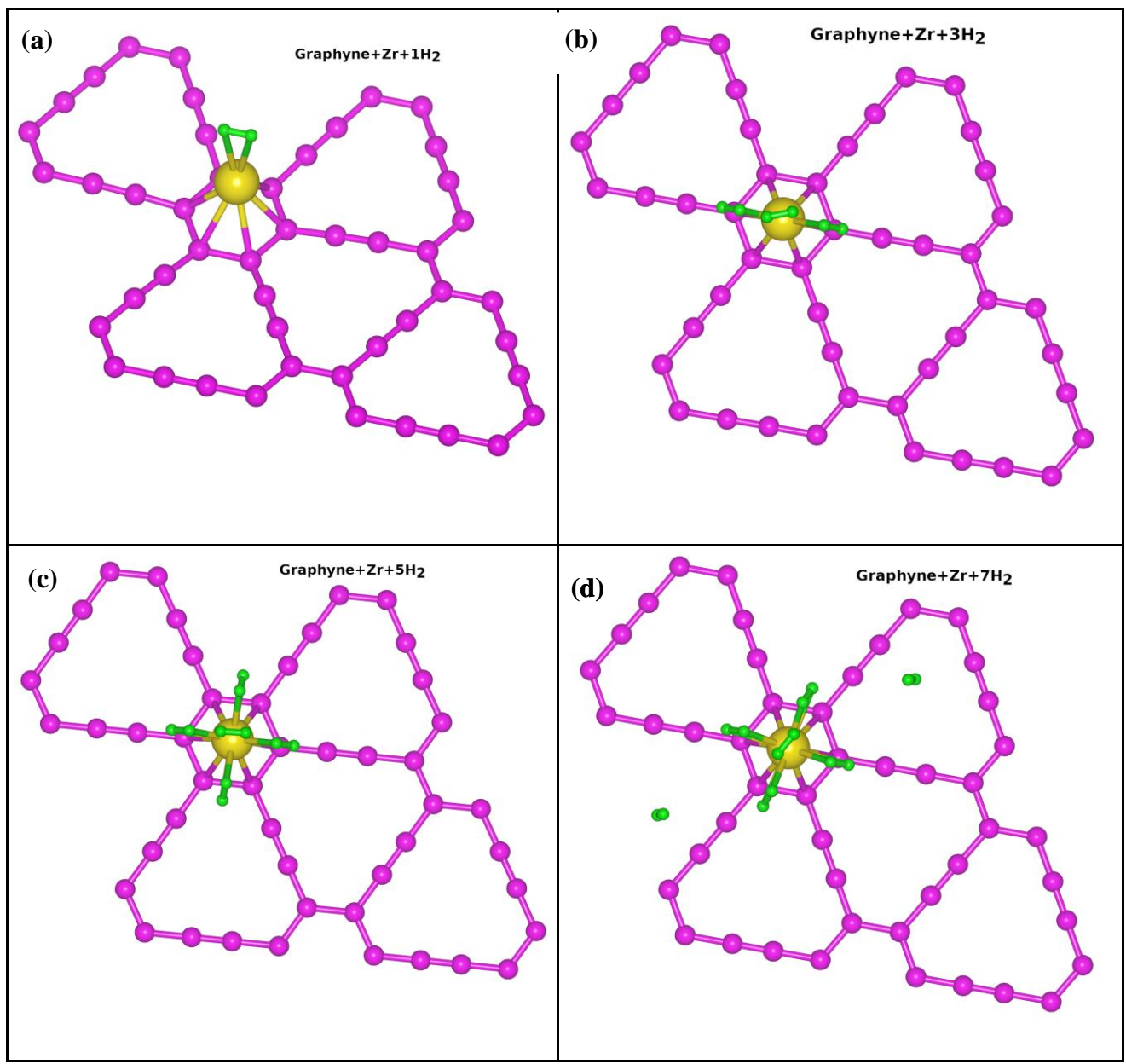

**Fig. 4**: DFT optimized structures of : **(a)** GY+Zr+1H$_2$; **(b)** GY+Zr+3H$_2$; **(c)** GY+Zr+5H$_2$; **(d)** GY+Zr+7H$_2$; Purple, golden and green colors represent carbon atom of graphyne sheet, doped Zirconium atom, and adsorbed hydrogen atoms, respectively.

**Table 2**: Comparison of gravimetric weight percentage, desorption temperature of hydrogen storage materials predicted by DFT and experimental reports.

| S.N. | System | Average desorption temperature(K) | H$_2$ wt% |
|---|---|---|---|
| 1. | Na+C60 [30] | 323 | ~9.5 |
| 2. | Y+SWCNT [37] | 396 | 6.1 |
| 3. | Li/Ca/Sc/Ti+graphyne nanotube [51] | -- | 4.82/5.08/4.88/4.76 |
| 4. | Sc + 6,6,12-graphyne nanotube [52] | -- | 5.4 |
| 5. | Ca+graphyne [52] | -- | 5.6 |
| 6. | Li/Na+boron-graphdiyne [55] | ~300 | 8.8/7.7 |

| 7. | Zr+graphyne (current system) | 7.95 |
| --- | --- | --- |
|    | **Experimental reports**     |      |
| 8. | $Pb_3Co$-NG/$Pb_3Co$-BG [45] | ~4.2/4.6 |
| 9. | Al/Ni/graphene [46]          | ~5.7 |
| 10.| Mg/$MgH_2$+Ni/GLM [47]       | >6.5 |

### 3.31 Computation of desorption temperature

The desorption temperature can be calculated by using von't Hoff equation [71]

$$T_d = \frac{E_b}{k_B}\left(\frac{\Delta S}{R} - lnP\right)^{-1} \tag{4}$$

In the equation above, $T_d$, $E_b$, P, $\Delta S$ are the desorption temperature, adsorption energy, ambient pressure, and the change in entropy for $H_2$ molecules during the transformation from the gas phase to the liquid phase respectively. Other quantities, $k_B$, R are Boltzmann, universal gas constants. After plugging all the numerical values in the equation, we obtain 574K value of the average desorption temperature.

### 3.32 Computation of gravimetric wt% of $H_2$

Since graphyne sheet is a 2D material, Zr and $H_2$ can be doped on its both sides, resulting in more gravimetric efficiency of $H_2$ storage. More Zr atoms per unit area need to be loaded on the graphyne sheet to achieve more volumetric efficiency. However, given the high cohesive energy of bulk Zr, the presence of many Zr atoms on the GY sheet may cause Zr-Zr clustering, resulting in a significant reduction in the system's hydrogen storage capacity. Considering this, many patterns of metal loading can be thought of. However, the most efficient one is to put Zr atoms on both sides of each hexagonal site. In this case, gravimetric wt% for one-sided and both-sided metal loading are 5.6 and 7.9, respectively Fig. 5.

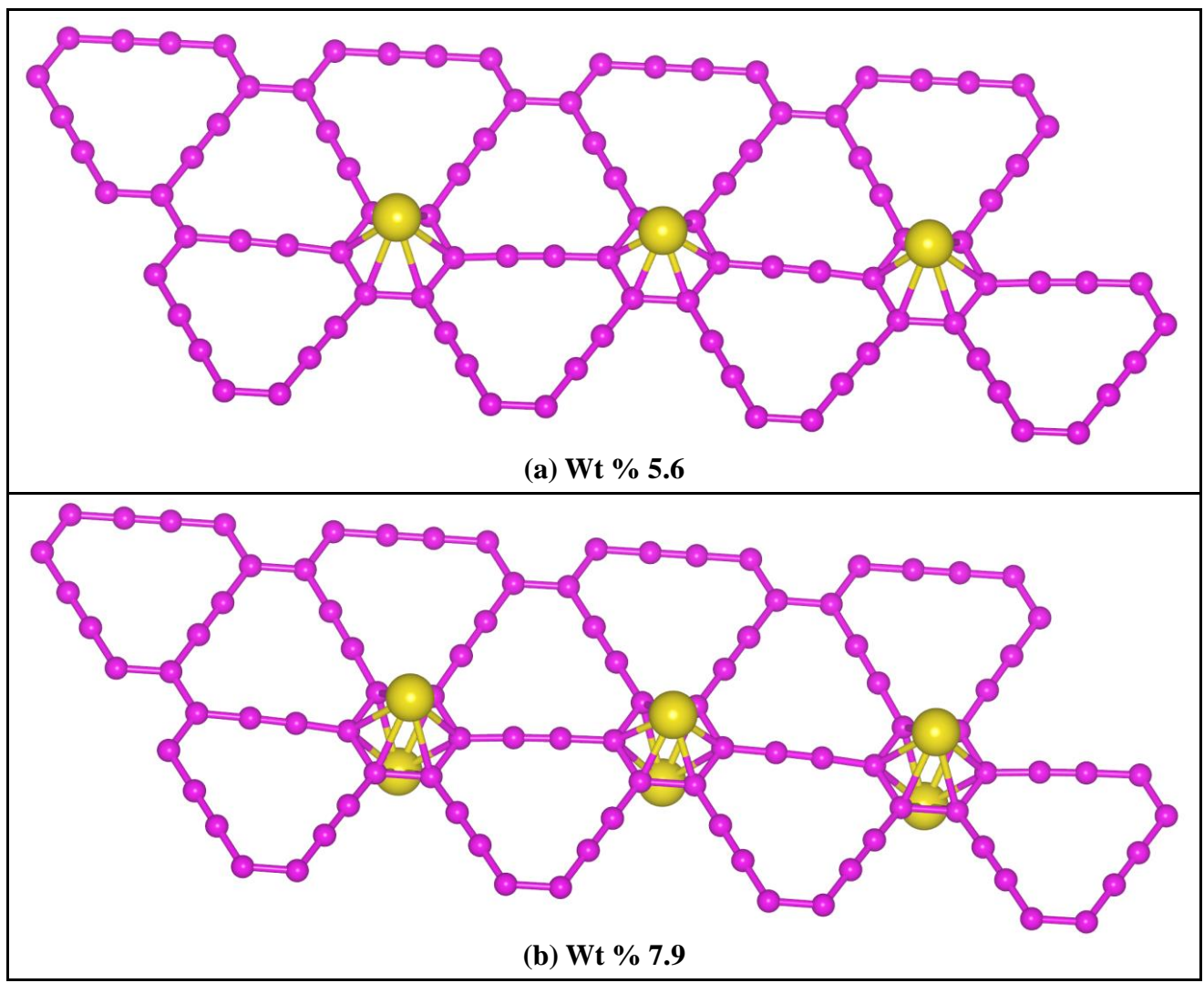

**Fig. 5:** Metal loading pattern on graphyne sheet for higher efficiency with their gravimetric wt% when considering (a) Zr decoration on one side of graphyne sheet (b) Zr decoration on both sides of graphyne sheet.

**3.4 The adsorption mechanism of hydrogens on Zr-decorated graphyne:** The adsorption energy of the first $H_2$ molecules is -0.432 eV (Table 1), which is greater than the weak interaction contributed by van der Waals interactions (physisorption). However, adsorbed $H_2$ molecules stay in molecular forms, and their binding energy is much smaller than the chemical binding range; therefore, the process cannot be chemisorption. We note that the enlarged bond lengths of the adsorbed $H_2$ molecules and their adsorption energies lie in between physisorption and chemisorption energy ranges. Hence, we conclude that interaction between $H_2$ and GY+Zr are Kubas type [72] in which the occupied 4d-electrons of Zr donate charge to the unoccupied anti-sigma bonds of $H_2$ molecules, which, in turn, back donate some charges to the unoccupied 4d-orbitals of the Zr atoms. And in this donation and back donation process, H2 molecules gain some charges resulting in their bond length elongation.

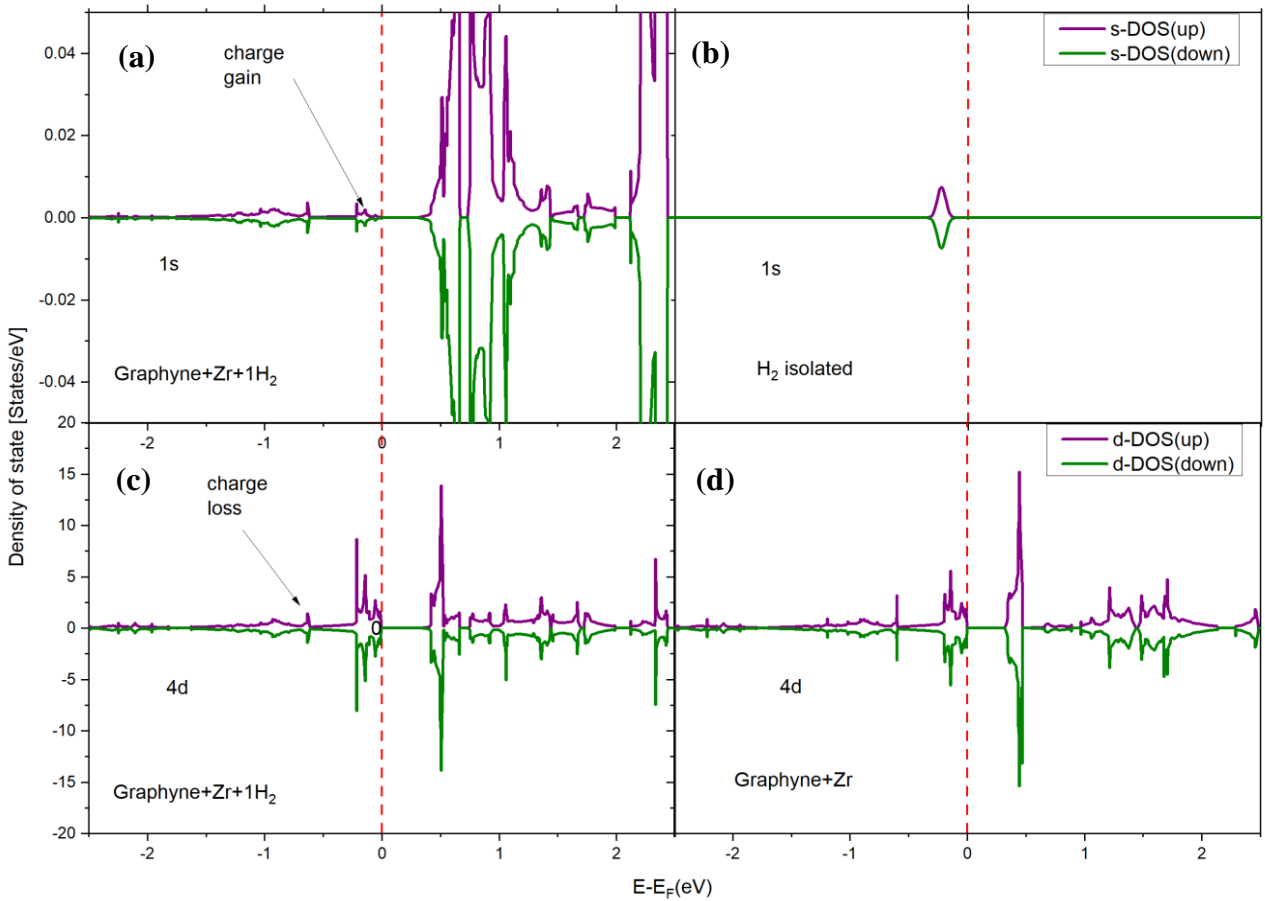

**Fig. 6:** Partial charge density of states of: **(a)** 1s orbital of hydrogen atom in GY+Zr+1H$_2$; **(b)** 1s orbital of hydrogen atom in isolated H$_2$ molecules; **(c)** 4d orbital of Zirconium atom in GY+Zr+1H$_2$; **(d)** 4d orbital of Zirconium atom in GY+Zr. Fermi energies are equated to zero.

**3.41 PDOS analysis:** To better understand the interaction between H$_2$ molecules and the GY+Zr system, we plotted H 1s and Zr 4d orbitals' PDOS before and after H$_2$ molecule attachment on Zr-decorated graphyne in Fig. 6. In the upper panel, it is clear that in the H 1s orbital of the GY+Zr+1H$_2$ system, the charge population has increased as compared to the isolated H$_2$ molecule. Hence, some net charge has been gained by the H$_2$ molecule in this interaction. This extra charge leads to the stretching of hydrogen bond length from 0.74 Å to 0.77 Å, but it is not strong enough to dissociate the attached H$_2$ molecule as in Kubas type of interaction. Similarly, from the lower panel of the figure, the charge on the valence-band side (i.e., occupied) of the 4d orbital has reduced when compared to that of 4d orbitals of isolated Zr atoms. Hence, there is a net charge loss in the 4d orbital of Zr atom during its interaction with the graphyne sheet. This charge transfer leads to the redistribution of charges in the sub-orbitals of the 4d orbital of Zr, which can be seen from PDOS of suborbitals of Zr's 4d orbital of isolated Zr and GY+Zr in the left and right panels of Fig. 7. From the figure, it is obvious that on the valence band side only three of the five 4d suborbital (in this case, $d_{xy}$, $d_{z^2-r^2}$, $d_{x^2-y^2}$) orbitals contribute to the PDOS, while after decorating Zr on graphyne, all five suborbital of the 4d of Zr contribute to the PDOS. As we load more hydrogen molecules on Zr atom, i.e., the system becomes GY+Zr+3H$_2$, GY+Zr+5H$_2$, etc., a similar charge reorganization will happen, but the net charge transfer will reduce with successive adsorptions. Hence, only a certain number of hydrogen molecules can be adsorbed at GY+Zr system. For our system, 7H$_2$ molecules can be adsorbed on each Zr at the hexa position. In Fig. 8, the orbital PDOS of the 4d orbital of Zr has been plotted for GY+Zr, GY+Zr+5H$_2$, GY+Zr+7H$_2$, showing that electrons populations at the Fermi level reduce as more and more hydrogen molecules are being adsorbed.

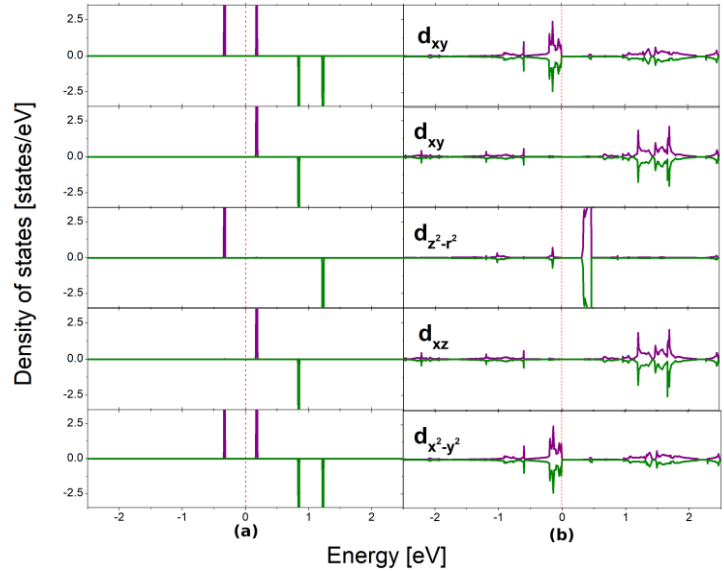

**Fig. 7:** Partial Density of States of suborbitals d$_{xy}$, d$_{yz}$, d$_{z2-r2}$, d$_{xz}$, and d$_{x2-y2}$ of the 4d atomic orbital of (a) isolated Zirconium, (b) Zirconium in GY+Zr system. So after doping Zr atom on graphyne, charges are redistributed in suborbital. Fermi-energy is equated to zero.

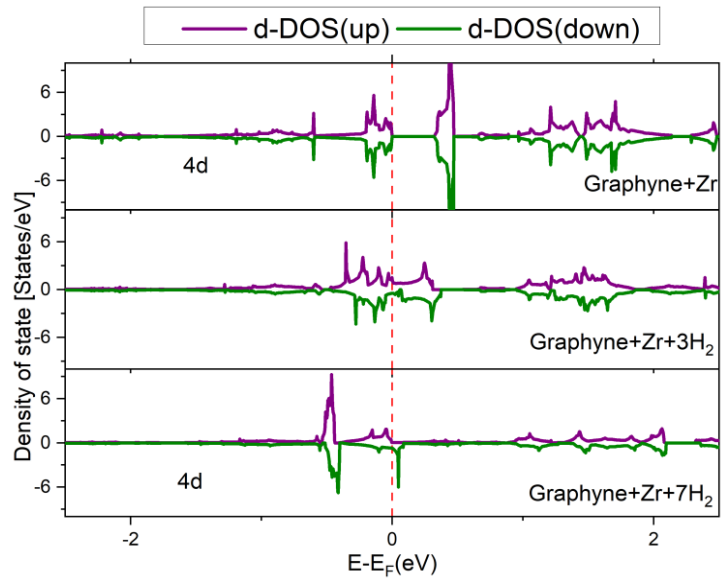

**Fig. 8:** PDOS of 4d orbital of Zr atom in: (a) GY+Zr (b)GY+Zr+3H2 ; (c) GY+Zr+7H2.

## 3.5 Practical feasibility:

**3.51 Energy barrier for diffusion:** The binding energy of Zr on graphyne (3.89 eV/atom) is less than the cohesive energy of bulk Zr (6.25 eV/atom [73]), so we have investigated the probability of metal-metal clustering in Zr doped graphyne by computing the energy barrier for movements of Zr atom from our considered system configuration, GY+Zr(h), to neighboring configuration, GY+Zr(t). The diffusion energy has been computed across the path from the hexagonal position surrounded by six carbon atoms to its neighboring trigonal position surrounded by 12 carbon atoms. We moved the Zr atom from the hexagonal position to the triangular position in small steps (see Fig. 9) and did the single-point calculation of total energy for each configuration. The energy difference ($E_i-E_o$) of each configuration with respect to ground energy ($E_o$) of hexagon position configuration, as a function of the moved distance ($R_i - R_o$), is plotted in Fig. 9. It shows that if Zr atom shifts to the nearest site, it faces an energy barrier of 4.05 eV. The energy barrier is greater than the thermal energy (0.095 eV) of Zr atom, so its diffusion is negligible even at the highest desorption temperature, avoiding the probability of metal-metal clustering.

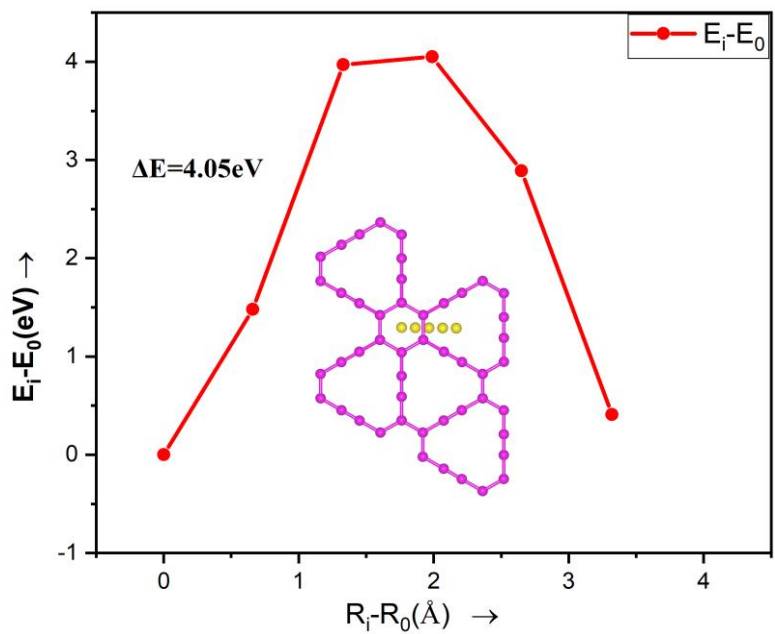

**Fig. 9**: Variation of energy difference ($E_i-E_o$) of different configurations with reference to Zr doped on the center of hexagon is plotted. The configurations are obtained by moving Zr atom from center of hexagon to center of triangle. Where ($R_i-R_o$) are displacement with respect to the initial state. Energy barrier, $\Delta E$, 4.05 eV provide stability.

## 3.52 AIMD calculations:

Since the DFT calculations are performed at absolute zero temperature, we need to check its stability at the room as well as the highest desorption temperatures. $H_2$ molecules should adsorb on GY+Zr system near room temperature and desorb at higher temperatures for an ideal solid-state hydrogen device. Hence, for durable and practical hydrogen storage, Zr atom needs to remain attached to graphyne sheet during the adsorption and desorption process, which requires the solidity of Zr+GY system from the room temperature to desorption temperature. The stability issues in the present case have been investigated by performing AIMD simulations in two steps: i) raise the temperature of the system with infinitesimal steps of 1 fs from 0 K to 300 K by using a microcanonical (NVE) ensemble for 5 ps ii) Keep the system at 300K in the canonical ensemble by equilibrating it with Nose-Hoover thermostats as demonstrated in Fig. 10(a). Our calculations indicate that the system remains stable, which is obvious from the bond length variation between Zr and its nearest carbon atom during the simulations plotted in Fig. 10(b). A negligible fluctuation of 8% for bond length (Zr-C) is noted. The same process of AIMD has been repeated for the highest desorption temperature. Again, the thermal stability of Zr decorated on both sides of graphyne is verified by AIMD in two different configurations at the highest desorption temperature, as plotted in Fig. 11. All finite-temperature simulations indicate that the system is fully stable up to the highest desorption temperature.

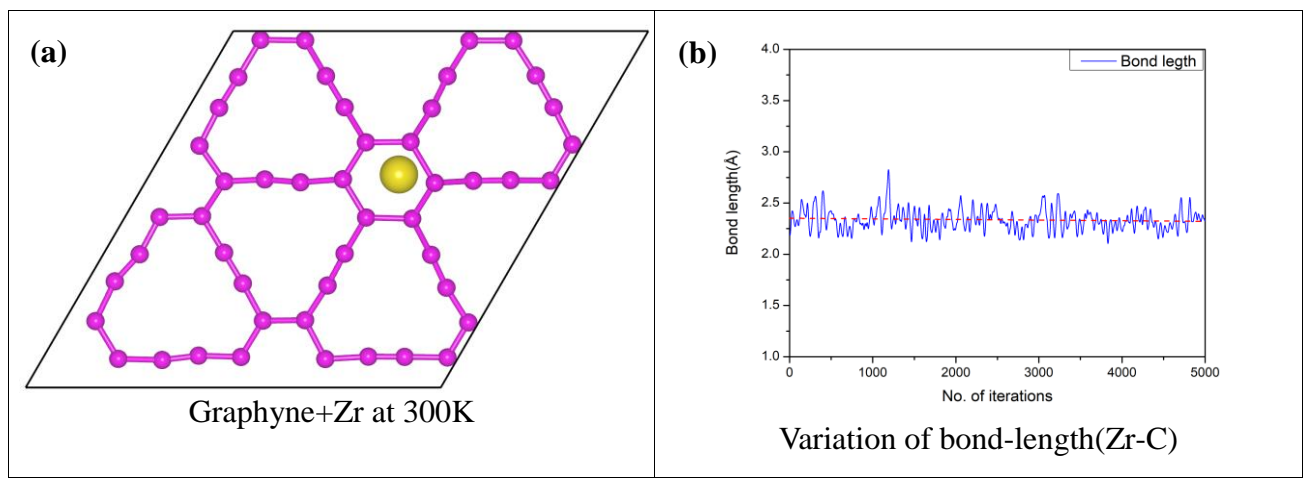

**Fig. 10**: (a) Molecular Dynamics snapshots of GY+Zr after 5 ps at 300 K. The purple and golden spheres are carbon atoms in the graphyne sheet and Zirconium atom, respectively. Zirconium atom has a minute displacement from its initial position. (b) Bond-length variation of Zr to nearest carbon-atom of graphyne sheet is ~8%.

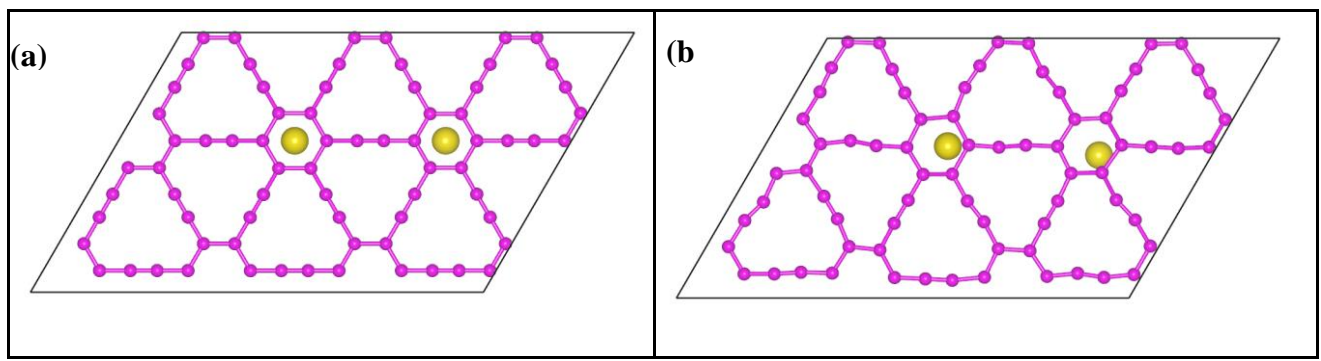

**Fig. 11:** Snapshot of GY+2Zr with Zr doped on opposite side of consecutive hexagon: (a) before AIMD; (b) after AIMD at the highest desorption temperature for 5 ps, i.e., first raise to 574 K for 5 ps and then simulate for 5 ps on constant temperature using Nose-Hoover thermostats in NVT ensemble.

## 4. Conclusion

Our first-principles DFT simulations predict that each Zr-decorated on the graphyne sheet can adsorb up to seven hydrogen molecules, with average binding energy and the desorption temperature of -0.44 eV and 574 K, respectively, which is quite suitable for fuel-cell applications. Zr interaction with graphyne sheet is due to charge transfer from 4d orbital of Zr to 2p orbital of C atoms of graphyne sheet. Also, the presence of acetylene bonds leads to strong binding at the triangular position. The interactions of hydrogen molecules and Zr are due to Kubas-type interactions. The high energy barrier between the center of the hexagon and its neighboring sites, triangles, may significantly reduce metal-metal clustering. The hydrogen gravimetric density of this system turns out to be 7.95 wt% which is greater than DoE's prescribed value of 6.5 wt%. Furthermore, the ab-initio molecular dynamic simulations show that the Zr-decorated graphyne is stable up to desorption temperature 574 K. Hence considering synthesized graphyne, Zr-decorated graphyne's wt%, stability, and no clustering, Zr-decorated graphyne might be a good choice for a hydrogen storage device. At the same time, fabrication of Zr-decorated graphyne samples with sufficient separations between Zr-Zr atoms are experimental challenges. We sincerely hope that this theoretical prediction of hydrogen storage encourages experimentalists to explore the hydrogen storage capacity of Zr-decorated graphyne for practical purposes.

ACKNOWLEDGEMENTS

MS would like to express his special thanks to CSIR for funding his Ph.D. research projects. MS also acknowledges the support of Spacetime High Performance Computing team and its facility at IIT Bombay. BC appreciates the assistance and encouragement from Dr. T. Shakuntala and Dr. Nandini Garg. BC acknowledges the insightful discussion with Dr. S.M. Yusuf and Dr. A. K. Mohanty.